\documentclass[superscriptaddress,twocolumn,pre,showpacs,preprintnumbers]{revtex4-1}
\usepackage{amsmath,amssymb}
\usepackage{epsfig,graphics,color,calc,graphicx}
\usepackage{mathrsfs}

\newcommand{\rr}[1]{{\normalfont\textrm{#1}}}

\newcommand{\bb}[1]{{\mathbb{#1}}}

\newcommand{\plow}{p}

\newlength{\pecettawidth}
\begin{document}
\title{Monte Carlo study of gating and selection in potassium channels}

\author{Daniele Andreucci}
\affiliation{Dipartimento di Scienze di Base e Applicate per 
             l'Ingegneria, Sapienza Universit\`a di Roma,
             via A.\ Scarpa 16, I--00161, Roma, Italy.}
\email{daniele.andreucci@sbai.uniroma1.it}

\author{Dario Bellaveglia}
\affiliation{Dipartimento di Scienze di Base e Applicate per 
             l'Ingegneria, Sapienza Universit\`a di Roma,
             via A.\ Scarpa 16, I--00161, Roma, Italy.}
\email{dario.bellaveglia@sbai.uniroma1.it}

\author{Emilio N.M.\ Cirillo}
\affiliation{Dipartimento di Scienze di Base e Applicate per 
             l'Ingegneria, Sapienza Universit\`a di Roma,
             via A.\ Scarpa 16, I--00161, Roma, Italy.}
\email{emilio.cirillo@uniroma1.it}

\author{Silvia Marconi}
\affiliation{Dipartimento di Scienze di Base e Applicate per 
             l'Ingegneria, Sapienza Universit\`a di Roma,
             via A.\ Scarpa 16, I--00161, Roma, Italy.}
\email{silvia.marconi@sbai.uniroma1.it}


\begin{abstract}
The study of 
selection and gating in potassium channels is a very important issue
in modern biology. Indeed such structures are known 
in essentially all types of cells in all organisms where they play 
many important functional roles. 
The mechanism of gating and selection of ionic species 
is not clearly understood. 
In this paper we study a model in which gating is obtained 
via an affinity--switching selectivity filter. 
We discuss the dependence of selectivity and efficiency on the 
cytosolic ionic concentration and on the typical pore open state 
duration.
We demonstrate that a simple modification of the way in 
which the selectivity filter is modeled yields larger channel efficiency.
\end{abstract}





\maketitle

\section{Introduction}
\label{s:introduzione}
\par\noindent
Potassium currents across nerve membranes have been
longly studied (see, for instance, \cite{HH,NS,Hille}
and  the reviews~\cite{GBOZ,VanDongen02,Miller,FH,RCM}).
It is now known that ionic channels selecting potassium currents 
are present in almost all types of cells in all organisms
and that they play many important and different functional roles.

An important improvement in experiments was the 
appearing of the 
\textit{patch clamp} technique, see for instance~\cite{SN},
which permitted the measurement
of the ionic current flowing through a single channel on the 
cell membrane. Different types of measurements, 
see for instance \cite{ADSEGHTM,ZM}, provide
a very detailed description of the behavior of potassium channels.
Less is known on their structure~\cite{Miller}; 
something has been inferred starting
from functional behavior and some has been deduced via direct inspection.

Despite the large variety of ionic channel types, they all
form selective pores in the cell membrane which open 
and close stochastically and, when in the open state, 
allow permeation of a selected ionic species (potassium in 
K$^+$--channels). Gating, i.e., their ability to open and close, and 
selectivity, i.e., their ability to allow the flux of a particular ionic 
species in the cytoplasm, 
are not yet completely understood. 
The way in which gating and selection are achieved can be different 
from channel to channel~\cite{Miller}.

On theoretical grounds this problem has been approached with a 
large variety of methods. 
Molecular dynamics studies \cite{Aqvist}
give a very detailed description of the phenomenon and 
their results can also be compared with structural experimental 
information, but they are usually not able to provide 
macroscopic currents estimates due to the too small involved time 
scale. 
A different approach is that of kinetic models
\cite{Miller02,Nelson,MP,Nelson02,Nelson03} which give very useful information 
since electro--physiological experiments are performed over time scales 
much longer than the atomic one. 
The drawback is represented by the extreme simplifications 
on the structure of the channel.

Models such as the ones in \cite{Nelson, Nelson02,Nelson03} 
describe to some extent 
the dynamics of ion permeation through the selectivity filter, see also 
\cite{MP}. In this paper, following \cite{VanDongen01}, we introduce 
a model where the channel is lumped to a two state stochastic point system. 
A remarkable 
feature of our model is the interaction between the dynamics of the 
ions inside the cell and that of the selectivity filter itself.

In this spirit, in~\cite{VanDongen01} a very simplified model for the pore 
behavior has been proposed. The main idea is that the pore can 
be in two states, \textit{high} and \textit{low}--affinity. 
In the former potassium ions can bind to molecules 
inside the pore, while smaller sodium ions cannot. In the latter, 
on the other hand, no ion can bind to the pore. 
We then have that, when the pore is in the low--affinity state, 
the two ionic species permeate in the same way. On the contrary, 
when the pore is the high--affinity state, sodium particles 
are reflected by the pore, while potassium ions bind to it and, 
eventually, when the pore comes back to the low-affinity state, dissociate
and possibly exit the cell.
This selectivity mechanism controls the potassium flux through the 
cell membrane and, hence, yields gating. 
In~\cite{VanDongen01} it was proven that it is possible to obtain 
gating via selection, but it was also noted that a 
pronounced reduction in the potassium flux is observed when the 
fraction of time the pore spends in the high--affinity state is large;
indeed in this case for long period of times no permeation is allowed.

Think for instance 
of the Kcv from chlorella virus PBCV--1~\cite{ADSEGHTM}, which have no long 
cytosolic regions able to close and open in response to membrane voltage. 
For this kind of channels it is attractive to think that gating 
is indeed realized via selection~\cite{VanDongen01}; in other words 
researchers were led to suppose that different current levels 
are indeed obtained through the selection mechanism. 

The basic motivation of this paper is  
studying the way in which the ion dynamics in cytosol 
affects the pore behavior and exploring the 
possibility to reduce the potassium flux loss due to 
selection. In our model we assume 
that the ionic species diffuse in the  
cytosol through independent random walks on a 
two--dimensional lattice and,  
for the potassium and sodium flux ratio,
we find results comparable to those in~\cite{VanDongen01}, 
where the motion of ions in cytosol was assumed to be uniform
(constant velocity).
As already noted in~\cite{VanDongen01}, 
good selection is obtained when the time fraction spent in the high--affinity 
state is chosen long enough. In this case, the potassium flux
through the channel is remarkably reduced. 
Our results suggest that, by modelling the cytosolic dynamics as a 
diffusive rather than uniform motion, 
the potassium flux loss is less important.
Moreover, we prove that it is possible
to compensate this effect just allowing the pore to stuff more 
than one particle at a time. For realistic concentrations
of potassium inside the cell it is seen that, allowing three (in 
rare occurencese four)
potassium ions to be packed in the pore, when it is in the 
high--affinity state, is enough to compensate the potassium 
flux reduction due to the selectivity filter.
This hypothesis is compatible with what is known about the structure 
of potassium channels \cite{Aqvist,Nelson}.

We also 
perform an extensive study of the dependence of 
fluxes on the two main physiological parameters, namely, 
the fraction of time the pore spends in the high--affinity state and the 
concentration of the ions in the cell. As it is easy to guess
the sodium flux is directly proportional to the 
ionic density in the cell. For potassium ions, on the other hand, 
this linear behavior is found only for small
values of the time fraction  
spent in the high--affinity state (in this case the 
two ionic species behave similarly), while for large values 
of such a time fraction a sub--linear behavior is found.
We studied the behavior
of fluxes as a function of the time fraction spent in the low--affinity 
state at fixed ionic concentration;
we found a sub--linear increasing behavior both for sodium and
potassium.

Since the pore is modeled in a very simple way 
(a two state random variable independent 
by the random walk dynamics inside the cell),
it is possible to study analytically the model. 
We perform an approximate study in the two--dimensional case
and we are able to explain the results found via the Monte Carlo  
simulation. 
The most interesting effect that is pointed out by this analysis 
is that the sub--linear behavior discussed above is due to a sort 
of depletion of the region close to the pore, which is observed 
when the typical time spent by the pore in the low--affinity state is 
large. 
The one dimensional model has been studied to some 
extent exactly in order to 
support our approximate two--dimensional analysis and to 
test the reliability of our Monte Carlo computation. 

The paper is organized as follows. In Section~\ref{s:modello} 
we introduce the model, describe the quantities that 
are measured with the Monte Carlo procedure, and give their
physical interpretation. 
Sections~\ref{s:risultati}--\ref{s:discussione} are devoted to the discussion 
of our numerical results: we first focus on the possibility of 
realizing gating 
via selection and on the way in which the potassium flux loss 
can be reduced; then we discuss the dependence of 
the fluxes on the physiological parameters of the model.
In Section~\ref{s:toy} we give an approximate analytical study of 
the two--dimensional model and in 
Section~\ref{s:conclusioni} we summarize our conclusions.
Appendices are devoted to the detailed algorithmic definition of the 
model and to the exact one--dimensional study.

\begin{figure*}
\begin{picture}(200,220)
\put(-100,-20)
{
\includegraphics[height=8cm,angle=0]{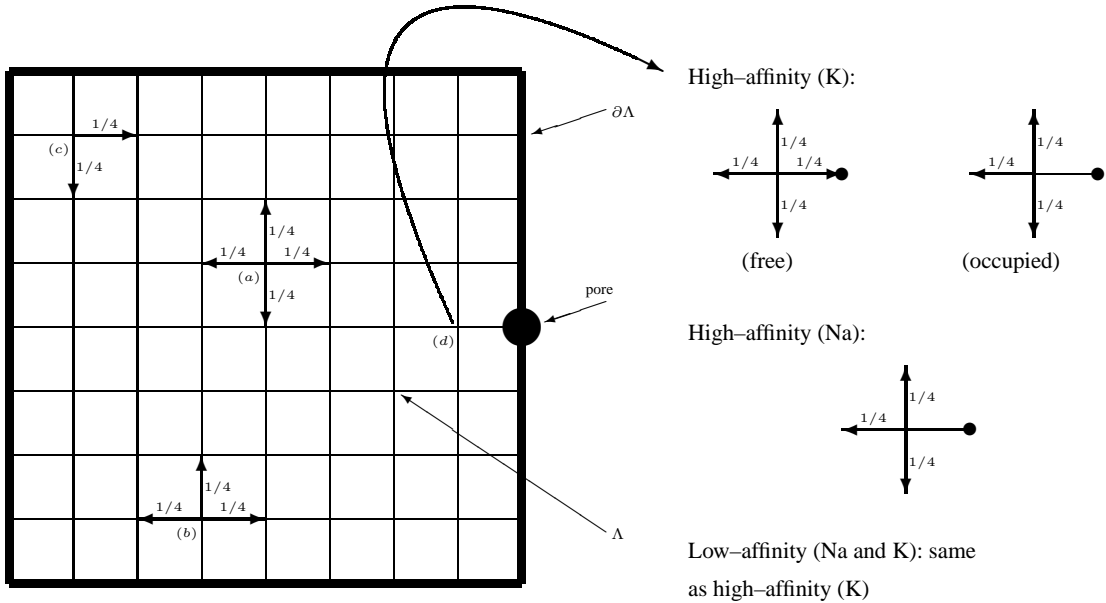}
}
\end{picture}  
\vskip 1. cm
\caption{On the left: thin line intersections represent the lattice
sites; intersections between thin and thick lines represent 
the boundary sites. The pore is in the middle of the 
right--hand boundary side.
(\textit{a}) Rules for a particle in the bulk: the particle jumps with 
uniform probability $1/4$ to one the four nearest neighboring sites.
(\textit{b}) Rules for a particle close to the boundary: 
the particle jumps with 
uniform probability $1/4$ to one the three nearest neighboring sites in 
the lattice; with probability $1/4$ the particle does not leave the site.
(\textit{c}) Rules for a particle in the corner: 
the particle jumps with 
uniform probability $1/4$ to one the two nearest neighboring sites in 
the lattice; with probability $1/2$ the particle does not leave the site.
(\textit{d}) The rules for a particle in the site neighboring the pore depend 
on the state of the pore and are 
explained in the picture on the right.
Pore in the high--affinity state (potassium): 
when the pore is free the particle jumps with 
uniform probability $1/4$ to one of the four nearest neighboring sites
(pore included); when the pore is occupied 
the particle jumps with 
uniform probability $1/4$ to one of the three nearest neighboring sites in 
the lattice (it cannot enter the pore) 
and with probability $1/4$ the particle does not leave the site.
Pore in the high--affinity state (sodium): 
the particle jumps with 
uniform probability $1/4$ to one of the three nearest neighboring sites in 
the lattice (it cannot enter the pore)
and with probability $1/4$ the particle does not leave the site.
Pore in the low--affinity state: same rule as that for the 
potassium particle faced to the pore in the high--affinity state;
it is worth remarking that the pore in the low--affinity state will 
never be occupied (see the description of the algorithm in the text). 
}
\label{f:reticolo}
\end{figure*}

\section{The model}
\label{s:modello}
\par\noindent
The model that we study in this paper is 
inspired to the two--dimensional one proposed in~\cite{VanDongen01}.
We give now a quick glance to the model and postpone the 
precise description of the algorithm to the appendix~\ref{s:bold}.
The cytosolic region of the cell is modeled via a 
finite two--dimensional 
square lattice $\Lambda$ with $L^2$ sites, where 
$L$ is an odd integer 
(see figure~\ref{f:reticolo} (left)).
We consider a two dimensional model for simplicity, but as it will 
be discussed in the sequel, the model is able to capture the main 
features of the real behavior.
Two ionic species perform 
independent symmetric random walks on the lattice with reflectivity 
conditions on the boundary $\partial\Lambda$ of the 
volume $\Lambda$. 
One of the site of the boundary is special and it is called \textit{pore};
its behavior, that will be described below, makes the potassium walkers 
not independent.
The number of potassium and sodium walkers are 
respectively denoted by $N_\rr{K}$ and $N_\rr{Na}$.
The fact that the  walkers are independent on $\Lambda$ 
means that the position 
of a particle does not affect the motion of the others, in particular 
no constraint to the number of particles on each site is prescribed.

The pore is modeled by the site in the middle of one of the four sides 
of the boundary $\partial\Lambda$.
Two states are allowed for the pore: \textit{high--affinity} 
and \textit{low--affinity}. The pore switches between the 
two states randomly; the probability that the pore is 
in the low--affinity state is denoted by 
$\plow$.
The behavior of the particles on the site neighboring the pore depends 
on the state of the pore itself as it is precisely stated in 
figure~\ref{f:reticolo}.
The idea is the following: if the pore site is in the low--affinity state, 
both potassium and sodium ions can enter it; when they enter the pore 
they immediately dissociate so that, with probability $1/2$, they 
reenter $\Lambda$ and with the same probability exit the system.
If the pore is in the high--affinity state, sodium ions 
are reflected by the pore site and potassium ions are reflected by 
the pore if it is occupied, while they can enter it if it is free.
In this last case the ion does not dissociate immediately, but it remains 
bounded to the pore until its state is changed to the low--affinity one.
As noted above, due to the pore rule, the sodium walkers are independent
while the potassium ones are not.

Whenever an ion exits the system a particle of the same species is put 
at random with uniform probability $1/L^2$ on one of the $L^2$ sites 
in $\Lambda$ so that the number of sodium and potassium ions 
in the system remains constant.

The model described above is implemented with the 
Monte Carlo algorithm which will be described in 
detail in the appendix~\ref{s:bold}. 
An \textit{iteration} or \textit{sweep} of the Monte Carlo is the 
collection of the steps that are performed at each time $t$.
The number of iterations that are performed in a simulation is 
$t_\rr{max}$.

\subsection{Measured quantities and goal}
\par\noindent
The particles that exit the system model the outgoing ionic flux. 
This is precisely what we want to measure. For a single 
numerical experiment we let 
$M_\rr{Na}(t)$ (resp.\  $M_\rr{K}(t)$)
be the number of sodium (resp.\ potassium) 
particles which exited the system in the time interval $[0,t]$.
Moreover, we let the sodium (resp.\ potassium) \textit{flux} 
$f_\rr{Na}(t)$ (resp.\  $f_\rr{K}(t)$)
at time $t$ be  
$M_\rr{Na}(t)/t$ (resp.\  $M_\rr{K}(t)/t$).
Since we defined the Monte Carlo algorithm in such a way that 
the number of sodium and potassium ions keep constant in the 
volume $\Lambda$, both 
$f_\rr{Na}(t)$ and $f_\rr{K}(t)$
approach a constant value for $t\to\infty$.
To estimate these limiting values, which are denoted by
$\bar f_\rr{Na}$ and $\bar f_\rr{K}$, 
we perform different numerical experiments with 
number of iterations 
$t_\rr{max}$ sufficiently 
large, we then average the experimental measures
$f_\rr{Na}(t_\rr{max})$ and $f_\rr{K}(t_\rr{max})$ and compute 
the associated statistical errors.

Our aim is to measure 
$\bar f_\rr{Na}$ and $\bar f_\rr{K}$ and to prove that 
the model provides selection in an efficient way. 
The two ionic species interact in the same way with the pore
in the low--affinity state,
see the rules in figure~\ref{f:reticolo};
hence, when $\plow$ is close to one, no difference 
should be observed in the potassium and sodium fluxes. 
A different behavior should be observed at smaller
$\plow$, indeed when the pore is in the high--affinity 
state the sodium flux is blocked while some residual potassium 
flux should be recorded.

We shall then compute the ratio 
$\bar f_\rr{K}/\bar f_\rr{Na}$ and prove that it becomes large for 
$\plow$ small. Moreover we shall discuss if this
selection mechanism is efficient, namely, we shall see that 
selection is provided with a potassium flux comparable with 
the flux measured when the pore is constantly in the 
low--affinity state. 
In particular in the sequel we shall discuss 
a slightly different model which will provide selection without
flux reduction.

\subsection{Choice of the parameters}
\par\noindent
We first describe how the physiological parameters of the 
model have been chosen.
We have considered
different values of the number of potassium ($N_\rr{K}$)
and sodium ($N_\rr{Na}$) particles. Although cytosol is typically
much richer in potassium than in sodium, in order to check 
the efficiency of the selectivity filter,
we have always taken $N_\rr{Na}=N_\rr{K}$.
In all the simulations discussed in this section we have chosen $L=101$, 
while we considered the cases
\begin{displaymath}
N_\rr{Na}=N_\rr{K}=100,1000,3000,5000,10000
\end{displaymath}
and
\begin{displaymath}
\begin{array}{rcl}
\plow
&\!\!=&\!\!
0.001,0.003,0.005,0.007,0.01,0.03,\\
&&\!\!
0.05,0.07,0.1,0.3,0.5,0.7,1
\\
\end{array}
\end{displaymath}
Note that the total time fraction 
the pore spends in the low--affinity state is essentially equal to 
$t_\rr{max}\plow$.

It is worth noting that, provided $L=101$,
the most physiologically reasonable choice for 
the number of potassium ions among those listed above seems to be 
$N_\rr{K}=100$. 
This is supported by the following very rough computation. The 
concentration of potassium ions in cytosol is approximatively equal 
to $c=150$~mM (millimolar), that is $c$ mole per cubic meter. The radius 
of the potassium ion is $r=1.52\times10^{-10}$~m; 
assuming that 
each site of the lattice can accommodate at most $n$ 
ions, we can ascribe
the volume $n(2r)^3$
to each site of the lattice.
The number of potassium ions that must be put on the 
lattice is then
\begin{displaymath}
\textrm{number of potassium ions}
=
cL^2n(2r)^3n_\rr{A}
=25.88n
\end{displaymath}
where $n_\rr{A}=6.02\times10^{23}$ is the Avogadro number, i.e., the 
number of ions in a mole. 
Assuming that reasonable values of $n$ ranges between 1 and 10, 
we have that realistic values of the number of potassium ions on the 
lattices ranges from 25 to 250.

The sole Monte Carlo parameter to be fixed 
is the number of sweeps $t_\rr{max}$.
As explained above it must be taken large 
enough in order to get a good estimate of the outgoing fluxes.
In all the cases that we considered it was sufficient to 
take $t_\rr{max}=10^7$.

\begin{figure}
\begin{picture}(200,180)
\put(-35,170)
{
\includegraphics[height=8.5cm,angle=-90]{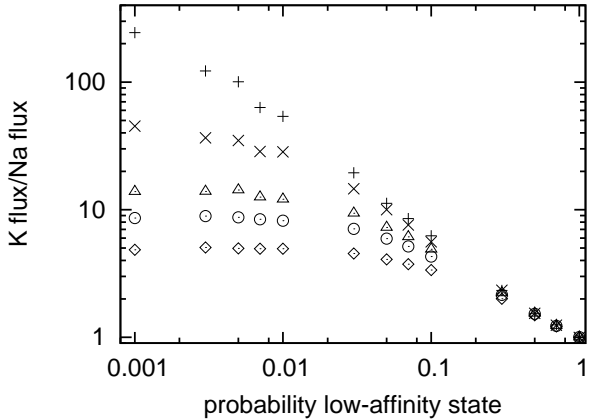}
}
\end{picture}  
\caption{Log--log 
plot of the ratio $\bar f_\rr{K}/\bar f_\rr{Na}$ as a function 
of the probability $\plow$.
Data refers to the cases 
$N_\rr{Na}=N_\rr{K}=10000 (\diamond), 5000 (\bigcirc),
                    3000 (\triangle),1000 (\times),100 (+)$.}
\label{f:ratio} 
\end{figure} 

\section{Gating and selection}
\label{s:risultati}
\par\noindent
As it has already been stated our first purpose is 
checking the behavior of the ratio $\bar f_\rr{K}/\bar f_\rr{Na}$ as a 
function of the probability $\plow$. 
Results are shown in figure~\ref{f:ratio}:
from the bottom to the top the ratio 
$\bar f_\rr{K}/\bar f_\rr{Na}$ for 
$N_\rr{Na}=N_\rr{K}=10000,5000,3000,1000,100$ has
been plotted as a function of $\plow$.
We have that the ratio between the potassium and the sodium 
outgoing fluxes increases when the probability for the 
pore to be in the low--affinity state is decreased. This is 
precisely the result in~\cite{VanDongen01}; in particular 
for $N_\rr{Na}=N_\rr{K}=100$ our data are not only qualitatively
but also quantitatively close to those in~\cite{VanDongen01}, although a 
different ion dynamics in the cytoplasm has been considered. 
As noted in~\cite{VanDongen01} it is then possible to imagine a 
potassium channel in which gating is realized via the selectivity filter.

The log--log plots in figure~\ref{f:ratio} shows a straight line 
behavior for $\plow$ large, say $\plow\ge0.08$. 
Note 
that the threshold where the linear behavior starts is smaller 
for smaller values of the ionic density in cytosol. 
This means that in this regime, 
say $\plow\ge0.08$, the ratio $\bar f_\rr{K}/\bar f_\rr{Na}$ 
can be well approximated by a power law
\begin{equation}
\label{ratio}
\frac{\bar f_\rr{K}}{\bar f_\rr{Na}}
=
a_1(\plow)^{-b_1}
\end{equation}
The coefficient $a_1$ and $b_1$ can be computed by fitting the 
data plotted in figure~\ref{f:ratio}; we obtain 
$a_1=0.863,0.904,0.974,1.014,1.072$ and 
$b_1=0.860,0.798,0.694,0.616,0.480$ 
respectively for 
$N_\rr{K}=N_\rr{Na}=100,1000,3000,5000,10000$.

It is worth noting that,
if the density of ions in the cytosolic region is not too high,
good selection ratios are reached at not too small $\plow$;
this is good news, since too low values of $\plow$
would imply very small ionic currents.
In particular, good results are found for
the physiologically reasonable 
parameters $N_\rr{Na}=N_\rr{K}=100$.
When the number of particles in the system is raised the 
selection effects are widely reduced. This is quite intuitive, indeed,
when the density is large, the average number of particles 
trying to enter the pore is large, as well; in other words the time 
between two consecutive attempts to enter the pore is small. If the 
typical duration of the intervals 
in which the pore stays in the high--affinity state 
is larger than this ``knocking" time, 
a lot of attempts to enter the pore will abort since the 
filter is occupied by another particle. It is then clear that the 
selection mechanisms does not work in the optimal regime.
In the sequel we shall propose a modified version of the 
model which will take care of this problem.

\begin{figure}
\begin{picture}(200,180)
\put(-35,170)
{
\includegraphics[height=8.5cm,angle=-90]{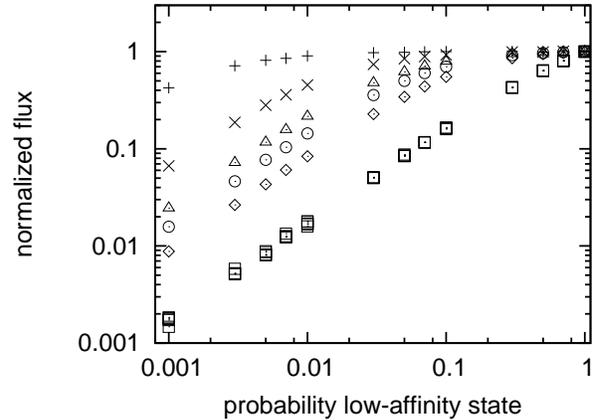}
}
\end{picture}  
\caption{Log--log 
plot of the normalized fluxes as functions of the low--affinity state 
probability $\plow$.
For the potassium 
the data in the picture refers to the cases 
$N_\rr{K}=10000 (\diamond), 5000 (\bigcirc),
                    3000 (\triangle),1000 (\times),100 (+)$;
for the sodium we used the symbol $\Box$ for the five cases
$N_\rr{Na}=10000, 5000, 3000, 1000, 100$.}
\label{f:normalizzo} 
\end{figure} 

We have seen that the algorithm provides good 
selection for $\plow$ small enough; at physiological concentrations, 
in order to get a ratio $\bar f_\rr{K}/\bar f_\rr{Na}$ larger than $100$, 
the low--affinity state probability $\plow$ must 
be chosen smaller than $0.01$. In this regime we expect a potassium
flux reduction with respect to the case $\plow=1$.
We estimate, now, this loss by plotting in figure~\ref{f:normalizzo}
the normalized potassium 
and sodium fluxes, namely, the ratio 
between fluxes at any $\plow$ with the 
corresponding fluxes at $\plow=1$.
We first note that the flux reduction is more effective for the sodium 
ions rather than for the potassium ones; this is obvious,
since the selectivity filter favors potassium particles. Moreover, 
in the case of sodium particles, the reduction does not depend 
on the number of particles on the lattice and this is due to the fact 
that all the sodium related measured quantities are directly proportional 
to the sodium density in cytosol.

Potassium flux, on the other hand, experiences a remarkable 
loss at small low--affinity state probability $\plow$. 
This reduction gets more and more important as the 
density is increased. 
At physiological potassium density 
the flux reduction is about the 40\% at very small low--affinity 
state probability. In respect to this point it seems 
that our algorithm is more efficient than the one 
proposed in~\cite{VanDongen01}, indeed, in that paper,
with a similar selection ability,
larger potassium flux losses were found.
In other words the importance of potassium flux loss
depends on how the cytosol dynamics is modeled, indeed in our model 
ions move inside the cell according to a symmetric random
walk, while in \cite{VanDongen01} they perform a uniform motion until
the boundary is reached. 
In the next section 
we propose a modified version of the model aiming to reduce 
potassium losses.

\section{Improving efficiency}
\par\noindent
As it has been discussed above, the model which has 
been proposed before 
is able to describe the gating attitude 
of a potassium channel via a suitable selectivity filter. 

The main problem with this approach is that the presence of the 
filter reduces the amount of the permeating potassium ions. To be more 
precise, in order to obtain a remarkable selection one has 
to assume that the time fraction in which the pore is in the low--affinity 
state is very small. From the data in figure~\ref{f:ratio}, it follows
that, at physiological concentrations, 
in order to get the ratio $\bar f_\rr{K}/\bar f_\rr{Na}$ larger or equal 
to 100 one has to assume $\plow\le0.1$.
As shown in figure~\ref{f:normalizzo}, at these values of the 
low--affinity state probability an important reduction of the 
potassium flux is observed. 

This phenomenon has a simple explanation: at small $\plow$ many 
attempts performed by K$^+$ ions to exit the cell abort since 
it often happens that 
the pore is in the high--affinity state and occupied by another 
potassium particle. 
This problem can be bypassed by assuming that 
more than one ion at time
can be accommodated in the pore.
We test this idea in the simplest possible way: we do not 
model the pore structure or any possible interaction between 
ions in the pore as it has been done in~\cite{VanDongen01} 
(this is obviously a very interesting question deserving 
future investigation). We simply assume that there is no 
upper bound to the number of ions that can be accommodated in the 
pore (see below).
In other words when the filter is in the high--affinity state, 
a potassium particle can enter it even if the pore is currently 
occupied by one or more potassium ions. 

This modified version of our model will assure no potassium flux reduction
since all the potassium ions trying to enter the pore in the high--affinity 
state will be accepted and then will be released when the state of the filter 
will switch to the low--affinity one. 

This model will not be physiologically reasonable 
if the typical number of potassium ions stuffed inside the pore 
is large. 
From what it is known on the structure of ionic channels
(see, for instance, \cite{Choe,Swartz,Aqvist}), a potassium 
channel is about 
$12$~\AA\ long and $2.5$~\AA\ in diameter. This means, recalling that 
the radius of a potassium ion is approximatively $1.5$~\AA, that 
at most four or five potassium particles 
can coexist inside 
the pore. We stress that in this simple model we do not take 
into account any interaction, for instance electrostatic repulsion,  
between ions in the pore. 
As we shall see, at physiological densities, i.e., $N_\rr{Na}=N_\rr{K}=100$
and for $\plow\ge0.01$, 
the average number of potassium ions in the pore will be very low 
and at most four particles (in very rare cases) will be simultaneously 
accommodated inside the pore. We shall then conclude that in this 
regime our simple model is reasonable on physiological grounds.

\begin{figure}
\begin{picture}(200,180)
\put(-35,170)
{
\includegraphics[height=8.5cm,angle=-90]{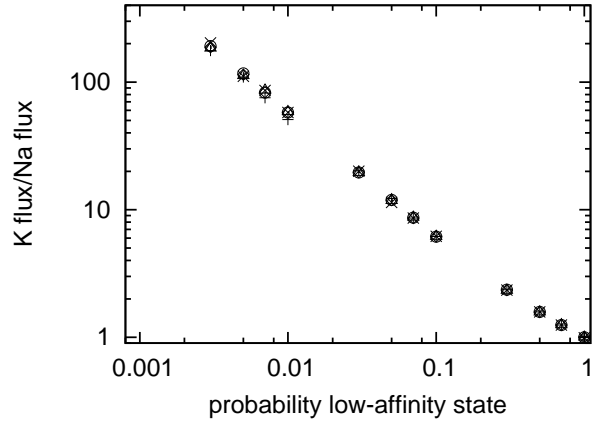}
}
\end{picture}  
\caption{Log--log 
plot of the ratio $\bar f_\rr{K}/\bar f_\rr{Na}$ as a function 
of the probability $\plow$ for the modified model; this picture 
should be compared with figure~\ref{f:ratio}.
Data refers to the cases 
$N_\rr{Na}=N_\rr{K}=10000 (\diamond), 5000 (\bigcirc),
                    3000 (\triangle),1000 (\times),100 (+)$.}
\label{f:ratio-inf} 
\end{figure} 

\begin{figure}
\begin{picture}(200,180)
\put(-35,170)
{
\includegraphics[height=8.5cm,angle=-90]{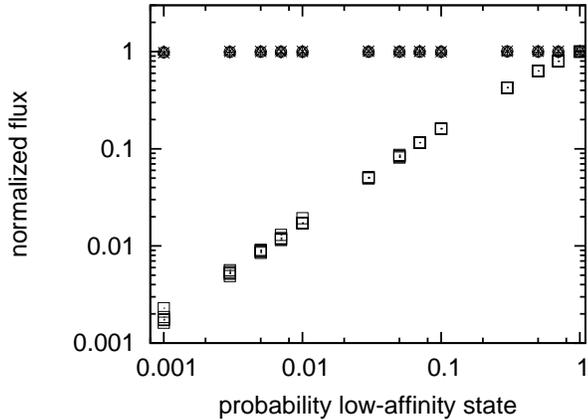}
}
\end{picture}  
\caption{Log--log 
plot of the normalized fluxes as functions of the low--affinity state 
probability $\plow$ for the modified model; this picture should 
be compared with figure~\ref{f:normalizzo}.
For the potassium 
the data in the picture refers to the cases 
$N_\rr{K}=10000 (\diamond), 5000 (\bigcirc),
                    3000 (\triangle),1000 (\times),100 (+)$;
for the sodium we used the symbol $\Box$ for the five cases
$N_\rr{Na}=10000, 5000, 3000, 1000, 100$.}
\label{f:normalizzo-inf} 
\end{figure} 

The results are shown in figures~\ref{f:ratio-inf} and 
\ref{f:normalizzo-inf}. By comparing figures~\ref{f:ratio-inf} 
and \ref{f:ratio}, it is clear that the selectivity attitude 
of the new model does not depend on the density of the 
ions in cytosol and that the modified 
model is more selective.
Figure~\ref{f:normalizzo-inf} is even more interesting, indeed 
it shows that 
the potassium flux does not depend on the 
low--affinity state probability.
In other words the model introduced in this section 
behaves precisely as we expected and, hence, it provides 
the solution to the potassium flux loss. 

\begin{figure*}
\begin{picture}(200,160)(130,0)
\put(0,130)
{
\includegraphics[height=5.cm,angle=-90]{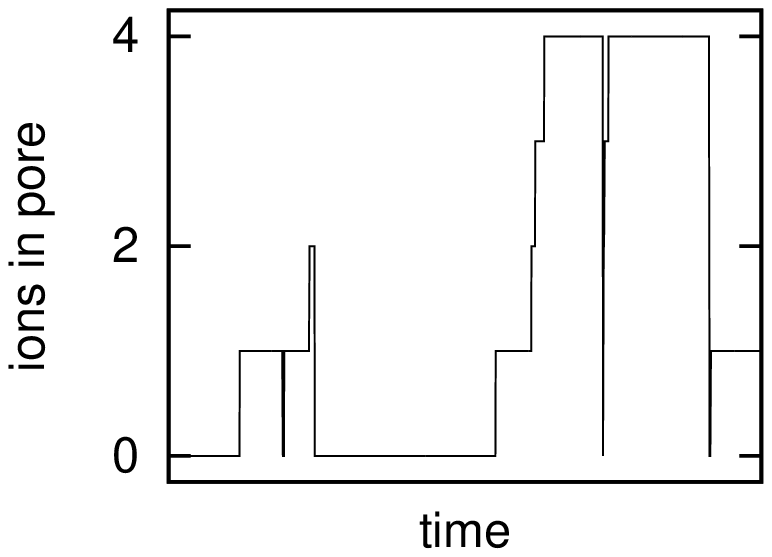}
}
\put(150,130)
{
\includegraphics[height=5.cm,angle=-90]{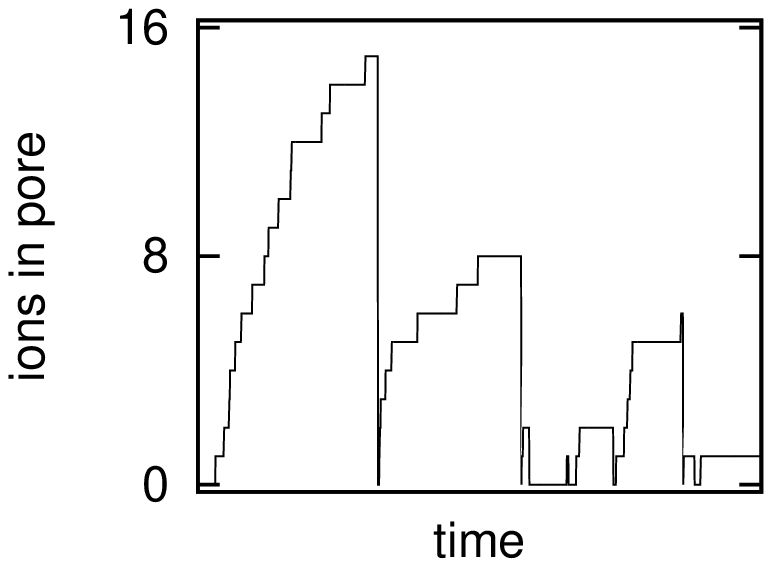}
}
\put(300,130)
{
\includegraphics[height=5.cm,angle=-90]{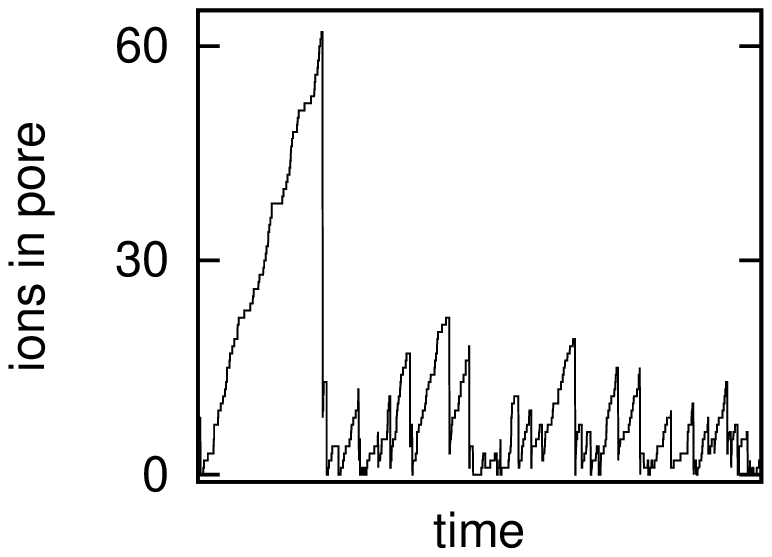}
}
\put(80,5){(a)}
\put(230,5){(b)}
\put(380,5){(c)}
\end{picture}  
\caption{Number of potassium ions accommodated inside the 
pore in an interval of time of few thousands sweeps out 
of $3\times10^7$ for the 
modified model. 
Parameters of the simulation: 
$\plow=0.01$,
$N_\rr{Na}=N_\rr{K}=100$ (a),
$N_\rr{Na}=N_\rr{K}=1000$ (b),
$N_\rr{Na}=N_\rr{K}=5000$ (c).
}
\label{f:accomodo} 
\end{figure*} 

It is important, now, to discuss the physiological reasonableness
of the model. In order to answer this question, which is 
very important in the present context, we have 
run a long simulation (about $3\times10^7$ sweeps) 
recording at each instant the 
number of potassium particles accommodated in the pore site. 
In the three cases $N_\rr{Na}=N_\rr{K}=100,1000,5000$ (see 
figure~\ref{f:accomodo} (a), (b), and (c)) 
the number of simultaneously stuffed particles is, along the
whole simulation, 
respectively smaller than $4$, $15$, and $62$.

It then follows that at physiological values of the cytosolic 
ion density, i.e., for $N_\rr{Na}=N_\rr{K}=100$ and for 
low--affinity state probability not too small ($\plow\ge0.01$), 
the model is completely reasonable. 

We can then conclude that a model of potassium channels is feasible
in which 
the gating attitude is obtained via a selectivity filter and such 
that the filter does not produce any potassium flux reduction.

\section{Discussion}
\label{s:discussione}
\par\noindent
We now discuss the behavior of the sodium and potassium flux
in the basic model as function of the two main 
physical parameters: the ionic cytosolic density and the 
time fraction in which the pore is in the low--affinity state (which, we 
recall, is 
essentially $\plow$).
In this section we shall simply describe 
the numerical results; the next section will be devoted 
to their physical interpretation via a simple analytical model.

It is difficult to have an intuition on the
potassium flux behavior, indeed the way in which 
the selectivity filter acts upon its current is not transparent at all.
Since the filter acts simply as a gate on 
sodium, it is rather natural to expect that 
the sodium flux will be directly proportional to the 
sodium density on the lattice; on the other hand 
the way in which it depends on the 
the low--affinity state probability, i.e., on the time fraction 
the filter spends in the low--affinity state, cannot 
be easily guessed a priori.

\begin{figure}
\begin{picture}(200,180)
\put(-35,170)
{
\includegraphics[height=8.5cm,angle=-90]{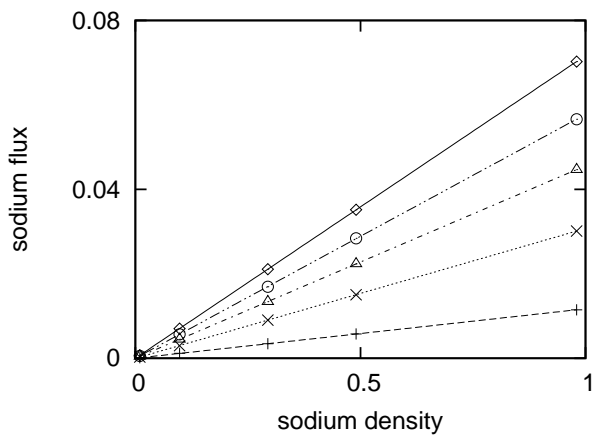}
}
\end{picture}  
\caption{Sodium flux $\bar f_\rr{Na}$ as a function of the 
sodium density $\varrho_\rr{Na}$.
Data in the picture refers to the cases 
$\plow=1 (\diamond), 0.7 (\bigcirc),
                    0.5 (\triangle), 0.3 (\times), 0.1 (+)$.
The other values of $\plow$ are not shown: the behavior is linear
with smaller and smaller slope (see table~\ref{t:fna-densita}).
Note that the behavior in the picture is much reasonable since 
the selectivity filter does not affect the sodium dynamics.
}
\label{f:fna-densita} 
\end{figure} 

\begin{figure}
\begin{picture}(200,180)
\put(-35,170)
{
\includegraphics[height=8.5cm,angle=-90]{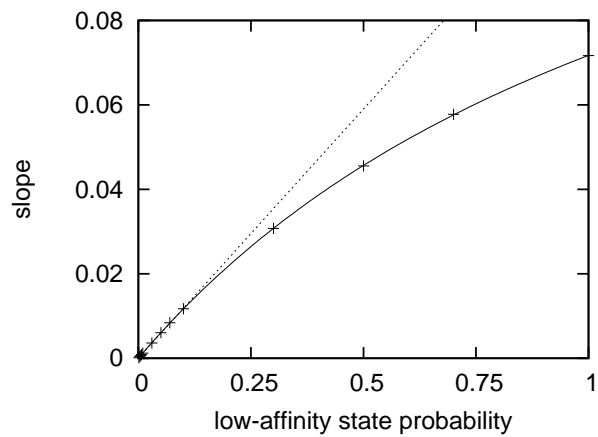}
}
\end{picture}  
\caption{Plot of the slope
coefficient $b_2$ introduced in equation (\ref{fna-densita}) as 
a function of the low--affinity state probability 
$\plow$.
The dotted straight line has been obtained by fitting the data with the 
function $b_2=a_3+b_3\plow$ on the interval $[0,0.1]$; 
the result of the fit is $a_3=0$ and $b_3=0.118$.
The solid line has been obtained by fitting the data with the 
function $b_2=\plow/(a'_3+b'_3\plow)$ on the whole 
data set, i.e., on the interval $[0,1]$; 
the result of the fit is $a'_3=7.96$ and $b'_3=5.98$;
note that the derivative in $0$ is $1/a'_3=0.1256\approx b_3$.
}
\label{f:slope} 
\end{figure} 

\begin{table}
\begin{center}
\begin{tabular}{c|c|c||c|c|c}
\hline\hline
$\plow$ & $a_2$ & $b_2$ &
$\plow$ & $a_2$ & $b_2$\\
\hline\hline
0.001 & 0 & 0.000131 &
0.07 & 0 & 0.008424\\
\hline
0.003 & 0 & 0.000376 &
0.1 & 0 & 0.011714\\
\hline
0.005 & 0 & 0.000626 &
0.3 & 0 & 0.030739\\
\hline
0.007 & 0 & 0.000877 &
0.5 & 0 & 0.045562\\
\hline
0.01 & 0 & 0.001220 &
0.7 & 0 & 0.057753\\
\hline
0.03 & 0 & 0.003608 &
1.0 & 0 & 0.071680\\
\hline
0.05 & 0 & 0.006034 &
&&\\
\hline\hline
\end{tabular}
\end{center}
\caption{Values of the parameters $a_2$ and $b_2$ introduced in 
equation (\ref{fna-densita}) and obtained by fitting the data plotted in 
figure~\ref{f:fna-densita}.}
\label{t:fna-densita} 
\end{table}

We discuss first how the sodium flux depends on concentration in 
cytosol. In figure~\ref{f:fna-densita} we have plotted 
the sodium flux $\bar f_\rr{Na}$ as a function of the sodium 
density $\varrho_\rr{Na}=N_\rr{Na}/L^2$ on the lattice.  
Only the curves corresponding to the values 
$\plow=1, 0.7, 0.5, 0.3, 0.1$ are shown; other values of the 
low--affinity state probability yield similar results.
As expected, the flux is directly proportional to the density 
with slope depending on the low--affinity state probability. 
We have fitted our data with the function 
\begin{equation}
\label{fna-densita}
\bar f_\rr{Na}=a_2+b_2\varrho_\rr{Na}
\end{equation}
and computed the parameters $a_2$ and $b_2$ for the different 
values of $\plow$; results are shown in the table~\ref{t:fna-densita}.
It is very interesting to remark that the slope 
coefficient $b_2$ depends linearly on $\plow$ in the 
interval $[0,0.1]$; for larger values of the low--affinity state 
probability the behavior departs from linearity. 
To illustrate this remark, in figure~\ref{f:slope} we have plotted 
the coefficient $b_2$ as a function of $\plow$ (we recall 
that the time fraction spent by the pore 
in the low--affinity state is, roughly speaking, equal to 
$\plow$).
The picture shows that for small low--affinity state probabilities 
the slope $b_2$ is directly proportional to $\plow$, 
that is to the time fraction
that the pore spends in the low--affinity state.
This result is 
consistent with a theorem proven in~\cite{AB} 
in a three--dimensional diffusion model. 

\begin{figure}
\begin{picture}(200,200)
\put(-35,170)
{
\includegraphics[height=8.5cm,angle=-90]{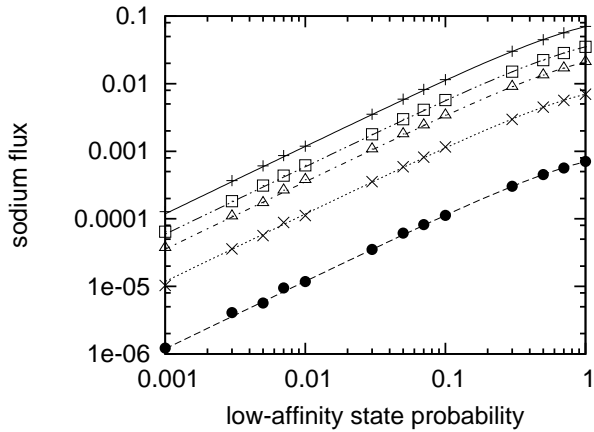}
}
\end{picture}  
\caption{Log--log plot of the sodium flux $\bar f_\rr{Na}$ as a function of 
the low--affinity state probability $\plow$.
Data refers to the cases 
$N_\rr{Na}=100 (\bullet), 1000 (\times), 
                    3000 (\triangle), 5000 (\Box), 10000 (+)$.
Lines are the graph of the function (\ref{fna-plow}) with the 
appropriate values of $\varrho_\rr{Na}$.
}
\label{f:fna-plow} 
\end{figure}

The predictions of (\ref{fna-densita}) with the coefficient $a_2,b_2$ 
chosen as in the caption of figure~\ref{f:slope}, that is 
\begin{equation}
\label{fna-plow}
\bar f_\rr{Na}=\varrho_\rr{Na}\frac{\plow}{a'_3+b'_3\plow}
\end{equation}
are compared with the experimental data in 
figure~\ref{f:fna-plow} and the matching is striking.

\begin{figure}
\begin{picture}(200,180)
\put(-35,170)
{
\includegraphics[height=8.5cm,angle=-90]{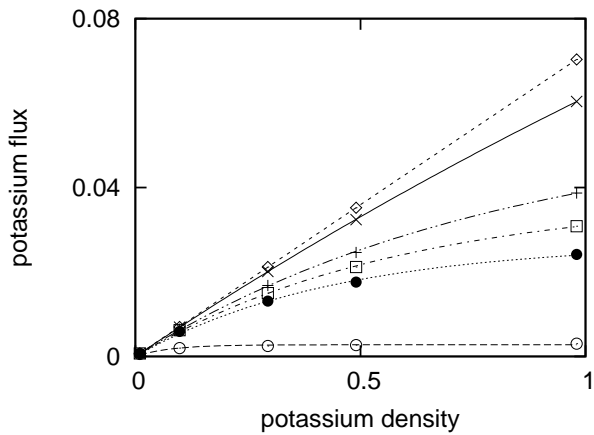}
}
\end{picture}  
\caption{Potassium flux $\bar f_\rr{K}$ as a function of the 
potassium density $\varrho_\rr{K}$.
Data in the picture refers to the cases 
$\plow=0.005 (\bigcirc), 0.05 (\bullet),
      0.07 (\Box), 0.1 (+), 0.3 (\times), 1.0 (\diamond)$.
The other values of $\plow$ are not shown: the behavior is similar. 
Lines are eye guides.
}
\label{f:fk-densita} 
\end{figure} 

We now come to the results related to the potassium flux. 
Potassium flux as a function of potassium density 
in the lattice bulk 
$\varrho_\rr{K}=N_\rr{K}/L^2$ 
has been plotted in figure~\ref{f:fk-densita}. In this case the 
behavior is approximatively linear for values of the 
low--affinity state probability $\plow$ close to one. But 
at smaller values of $\plow$ the graph remarkably 
departs from linearity. In other words, when the time fraction spent 
by the pore in the low--affinity state is large, the potassium flux is,
with good approximation, directly proportional to its density in 
the bulk. On the other hand, when the time fraction 
spent by the filter in 
the low--affinity state is smaller, the potassium flux 
exhibits a sub--linear behavior with density.

\begin{figure}
\begin{picture}(200,200)
\put(-35,170)
{
\includegraphics[height=8.5cm,angle=-90]{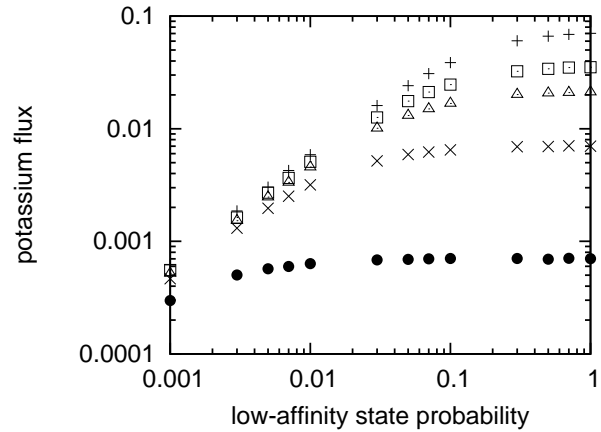}
}
\end{picture}  
\caption{Log--log plot of the potassium flux $\bar f_\rr{K}$ as a function of 
the low--affinity state probability $\plow$.
Data refers to the cases 
$N_\rr{K}=100 (\bullet), 1000 (\times), 
                    3000 (\triangle), 5000 (\Box), 10000 (+)$.
}
\label{f:fk-plow} 
\end{figure} 

Finally, in figure~\ref{f:fk-plow} 
the potassium flux as a function of the low--affinity state probability 
$\plow$ is reported.

\section{Analytical study}
\label{s:toy}
\par\noindent
In this section we study analytically the basic 
model introduced in Section~\ref{s:modello} in order to give a physical 
interpretation of the behavior of sodium and potassium flux 
on the species density and on the low--affinity state probability. 
In our model the 
pore is modeled in a very simple fashion, 
indeed it is just a two state Bernoulli process;
the main difficulty is, obviously, the interaction between 
the random walk inside the volume $\Lambda$ and the pore 
itself. We consider the 
stationary state reached by a walker and denote by 
$q_\rr{K}$ (resp.\ $q_\rr{Na}$)
the probability  
for a potassium (resp.\ sodium) 
ion to occupy the 
site $x$ of $\Lambda$ neighboring the pore.

\begin{figure}
\begin{picture}(200,200)
\put(-35,170)
{
\includegraphics[height=8.5cm,angle=-90]{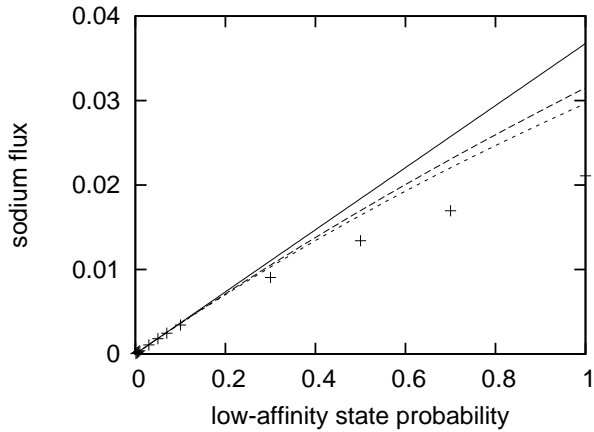}
}
\end{picture}  
\caption{Sodium flux as function of the low--affinity state probability.
Pluses are the Monte Carlo results for $N_\rr{Na}=3000$.
The solid, long dashed, and short dashed 
lines are the corresponding graphs of the 
functions 
$f^{(0)}_\rr{Na}(\plow)$,
$f^{(1)}_\rr{Na}(\plow)$
and $f^{(2)}_\rr{Na}(\plow)$ respectively.
}
\label{f:fnanew} 
\end{figure} 

Since sodium particles can exit the system only when 
the pore is in the low--affinity state, 
the sodium flux is simply given by
\begin{equation}
\label{flussona}
f_\rr{Na}=\frac{1}{8}N_\rr{Na}q_\rr{Na}\plow
\end{equation}
where, we recall, 
$N_\rr{Na}$ is the number of sodium ions in the volume $\Lambda$.
Indeed,
the probability for a sodium ion to be on the site $x$ neighboring 
the pore is $q_\rr{Na}$, 
the probability for such a particle 
to jump to the pore itself is $1/4$ and, once in the pore, 
the probability to really exiting the system is 
$1/2$.

When dealing with potassium,  
one has to take into account that, when the pore 
switches to the low--affinity state, the possibly
trapped potassium ion
is released with probability one half. 
We then have for the potassium flux 
\begin{equation}
\label{flussok}
f_\rr{K}
=
\frac{1}{8}N_\rr{K}q_\rr{k}\plow
+
\frac{1}{2}\plow r
\end{equation}
where we have denoted by $r$ the stationary probability for the pore 
to be occupied by one of the $N_\rr{K}$ potassium particles.

The problem, now, is that of computing the stationary probabilities
$q_\rr{Na},q_\rr{K}$, and $r$. The following study
is related to the two--dimensional model discussed in 
the above sections; in the appendix~\ref{s:uni}
we address the same problem in the one dimensional case, which 
is an interesting case, since some of the involved computations 
can be performed exactly.
The related results will both support our two--dimensional 
approximate discussion and confirm the reliability of our 
Monte Carlo measurements.

We first consider the sodium case.
An obvious guess is that the probability
$q_\rr{Na}$ 
is $q^{(0)}_\rr{Na}=1/|\Lambda|$, 
i.e., it is uniform throughout the system.
In this way we get 
\begin{equation}
\label{fsodio}
f^{(0)}_\rr{Na}
=
\frac{1}{8}N_\rr{Na}\frac{1}{L^2}\plow
\end{equation}
Note that this simple model provides the estimate $8$ 
for the constant $a'_3=7.96$, obtained by fitting the data 
in figure~\ref{f:slope},
and $0$ for the constant $b'_3=5.98$.
The comparison between the Monte Carlo data and 
the theoretical prediction 
$f^{(0)}_\rr{Na}$ is good for small values 
of the low--affinity state probability, while (see figure~\ref{f:fnanew}),
the experimental data departs from the predicted behavior for large
$\plow$.

\begin{figure}
\begin{picture}(200,200)
\put(-35,170)
{
\includegraphics[height=8.5cm,angle=-90]{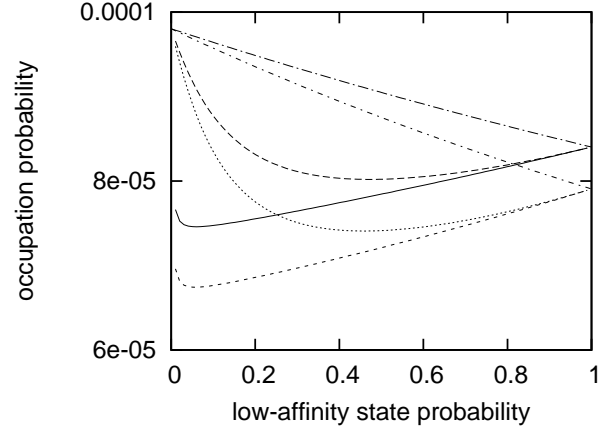}
}
\end{picture}  
\caption{Referring to the left side of the picture, 
from the top to the bottom 
the graphs of the functions
$q^{(1)}_\rr{Na}(\plow)$,
$q^{(2)}_\rr{Na}(\plow)$,
$q^{(1)}_\rr{K}(\plow)$ and 
$q^{(2)}_\rr{K}(\plow)$ for $N_\rr{K}=10000$,
$q^{(1)}_\rr{K}(\plow)$ 
and
$q^{(2)}_\rr{K}(\plow)$ for $N_\rr{K}=100$,
are plotted.}
\label{f:foccu} 
\end{figure}

This behavior is not surprising: at large $\plow$ the outgoing 
flux is so important that one has to expect a sort of depletion 
of the region close to the pore. 
To estimate this effect we develop a sort of ``mean field" argument: we imagine 
that in all the sites of $\Lambda$ different from the one 
neighboring the pore the stationary probability $q_{\rr{Na},0}$ is uniform.
To get the stationary probability $q_\rr{Na}$ in the site $x$
close to the pore 
we make the following probability balance
\begin{displaymath} 
q_\rr{Na}
=
\frac{3}{4}q_{\rr{Na},0}
+\frac{1}{4}(1-\plow)q_\rr{Na}
+\frac{1}{4}\frac{1}{2}\plow q_\rr{Na}
+\frac{1}{4}\frac{1}{2L^2}\plow q_\rr{Na}
\end{displaymath} 
where, on the right hand side, 
the first term is the probability that the particle moves 
from $\Lambda\setminus\{x\}$ to $x$, 
whereas the other terms 
take into account the 
probability that the particle in $x$ remains in $x$.
The above equation, together with the obvious normalization 
condition, 
\begin{displaymath}
(L^2-1)q_{\rr{Na},0}+q_\rr{Na}=1
\end{displaymath} 
yields
\begin{displaymath}
q^{(1)}_\rr{Na}=
\frac{1}{L^2+\plow(L^2-1)^2/(6L^2)}
\end{displaymath}
for the occupation probability and, hence,  
\begin{displaymath}
f^{(1)}_\rr{Na}=
\frac{1}{8}
N_\rr{Na}
\frac{\plow}{L^2+\plow(L^2-1)^2/(6L^2)}
\end{displaymath}
for the sodium outgoing flux.
This approximation provides the estimate $8$ 
for the constant $a'_3=7.96$, obtained by fitting the data 
in figure~\ref{f:slope}, 
and $4(L^2-1)^2/(3L^4)=1.33$ for the constant $b'_3=5.98$.
The graph of the occupation probability is plotted in 
figure~\ref{f:foccu}; the guessed depletion effect is 
found and the occupation probability decreases when 
$\plow$ is increased.

For $\plow$ small and $L$ large, 
$f^{(1)}_\rr{Na}\approx f^{(0)}_\rr{Na}$; 
this is reasonable, since in this limit the depletion effect 
is negligible.
As it is shown in figure~\ref{f:fnanew} the function $f^{(1)}_\rr{Na}$ 
is a quite good approximation of the experimental data, which, as we 
have seen above, are perfectly fitted by the function $\bar{f}_\rr{Na}$ 
given in equation (\ref{fna-plow}).

A more detailed description of the dynamics close to the pore 
can yield a better estimate of the depletion effect. 
Consider one of the $N_\rr{Na}$ walkers on $\Lambda$ and 
let, as we did before, $q_\rr{Na}$ be the probability that 
it occupies the site 
$x$ in $\Lambda$, neighboring the pore. 
Say that the boundary is along the north--south direction and that 
the pore is to the east of the site $x$.
Let $q_{\rr{Na},1}$ be the probability that the particle is on 
the site to the north of $x$ and that it is equal to the probability that 
the walker occupies the site to the south. 
Let $q_{\rr{Na},2}$ be the probability that the particle is on 
the site to the west of $x$.
Assume that for the remaining $L^2-4$ sites of the lattice the 
occupation probability is $q_{\rr{Na},0}$.

As before it is not difficult to write the following 
probability balance equations
\begin{displaymath}
\left\{
\begin{array}{l}
q_{\rr{Na},1}
=
2q_{\rr{Na},0}/4
+
q_{\rr{Na}}/4
+
q_{\rr{Na},1}/4
\\
q_{\rr{Na},2}
=
3q_{\rr{Na},0}/4
+
q_{\rr{Na}}/4
\\
q_{\rr{Na}}
=
q_{\rr{Na},2}/4
+
2q_{\rr{Na},1}/4
+
(1-\plow)q_{\rr{Na}}/4
+
\plow q_{\rr{Na}}/8
\\
\end{array}
\right.
\end{displaymath}
where we have neglected the terms proportional to $1/L^2$.
The above equations, 
together with the obvious normalization 
condition 
\begin{displaymath}
(L^2-4)q_{\rr{Na},0}
+
q_{\rr{Na},2}
+
2q_{\rr{Na},1}
+
q_\rr{Na}=1
\end{displaymath} 
yield
\begin{displaymath}
q^{(2)}_\rr{Na}=
\frac{12/23}{1+12(1+6\plow/25)(L^2-23/12)/23}
\end{displaymath}
for the occupation probability and, hence, 
\begin{displaymath}
f^{(2)}_\rr{Na}=
\frac{1}{8}
N_\rr{Na}
\frac{1}{L^2+6\plow(L^2-23/12)/25}
\plow
\end{displaymath}
for the sodium outgoing flux.
This approximation provides the estimate $8$ 
for the constant $a'_3=7.96$, obtained by fitting the data 
in figure~\ref{f:slope}, 
and $48(1-23/(12L^2))/25=1.92$ for the constant $b'_3=5.98$.
Note that for $\plow$ small and $L$ large, 
$f^{(2)}_\rr{Na}\approx f^{(0)}_\rr{Na}$.
As it is shown in figure~\ref{f:fnanew} the function $f^{(2)}_\rr{Na}$ 
is a better approximation of the experimental data with respect to 
$f^{(1)}_\rr{Na}$ and $f^{(0)}_\rr{Na}$. 

The graph of the occupation probability $q^{(2)}_\rr{Na}(\plow)$
is plotted in 
figure~\ref{f:foccu}; the depletion effect is 
more pronounced with respect to that described by 
the approximation $q^{(1)}_\rr{Na}(\plow)$.

\begin{figure}
\begin{picture}(200,200)
\put(-35,170)
{
\includegraphics[height=8.5cm,angle=-90]{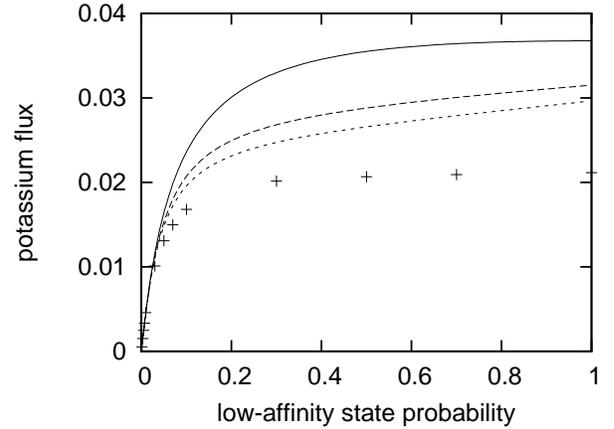}
}
\end{picture}  
\caption{Potassium flux as function of the low--affinity state probability.
Pluses are the Monte Carlo results for $N_\rr{K}=3000$.
The solid, broken (large), broken (small),
dotted lines are the corresponding graphs of the 
functions 
$f^{(0)}_\rr{K}(\plow)$,
$f^{(1)}_\rr{K}(\plow)$,
and
$f^{(2)}_\rr{K}(\plow)$
respectively.
}
\label{f:fknew} 
\end{figure} 

We now come to the computation of the potassium flux. 
We consider a potassium walker and we note that for such 
a particle the random walk is performed in $\Lambda\cup\{\textrm{pore}\}$.
We assume that in the stationary state the probability that the potassium
walker is in the pore is $r/N_\rr{K}$. 
Thus in the stationary state 
\begin{displaymath}
\frac{r}{N_\rr{K}}
=
\frac{1}{4}(1-\plow)(1-r)q_\rr{K}+(1-\plow)\frac{r}{N_\rr{K}}
\end{displaymath}
where, we recall, $q_\rr{K}$ is the probability that the potassium 
ion occupies the site $x$ in $\Lambda$ neighboring the pore.
Indeed, by stationarity, the probability that the particle is in the 
pore is equal
to the probability that the pore is empty in the high--affinity state 
and that the particle jumps from $x$ to it plus the probability 
that the pore is occupied and in the high affinity state.
From the equation above we easily get 
\begin{equation}
\label{erre}
r
=
\frac{1}{4}
N_\rr{K}q_\rr{K}(1-\plow)
\frac{1}{\plow+N_\rr{K}q_\rr{K}(1-\plow)/4}
\end{equation}

Then we have to estimate $q_\rr{K}$. 
As we have done in the case of sodium we
first consider the stationary occupation probability
uniform inside
the volume $\Lambda$ and equal to $q^{(0)}_\rr{K}=1/L^2$. The 
corresponding 
probability $r^{(0)}$ for the pore to be occupied 
is given by the equation (\ref{erre}) and the corresponding
potassium 
flux $f^{(0)}_\rr{K}$, given by (\ref{flussok}),
is plotted in figure~\ref{f:fknew}.
The comparison with the experimental data is very 
good at small low--affinity probability $\plow$, while 
the experimental results depart from the predicted behavior at 
large $\plow$.

As before this is due to the depletion of the region close to 
the pore. We repeat the same computation described in detail 
for the sodium particles. Here we just give the main formulas: 
the first approximation consists in the probability balance 
equation
\begin{displaymath}
\begin{array}{l}
{\displaystyle
 q_\rr{K}
 =
 \frac{3}{4}q_{\rr{K},0}
 +\frac{1}{8}q_\rr{K} \plow
 +\frac{1}{8L^2}q_\rr{K} \plow
 +\frac{1}{4}q_\rr{K} (1-\plow)r
 \vphantom{\Bigg[}
}
\\
{\displaystyle
 \phantom{ q_\rr{K} = }
 +\frac{1}{2L^2}\plow\frac{r}{N_\rr{K}}
 +\frac{1}{2}\plow\frac{r}{N_\rr{K}}
}
\\
\end{array}
\end{displaymath}
By using the obvious normalization condition 
$(L^2-1)q_{\rr{K},0}+q_\rr{K}+r/N_\rr{K}=1$, 
by neglecting all the terms proportional to $1/L^2$ and to $1/N_\rr{K}$, 
and by recalling equation (\ref{erre}), we get 
\begin{displaymath}
r^{(1)}
=
\frac{AD+B\plow-\sqrt{(AD+B\plow)^2-4ACD\plow}}{2C\plow}
\end{displaymath}
and 
\begin{displaymath}
q^{(1)}_\rr{K}
=
\frac{A}{B-Cr^{(1)}}
\end{displaymath}
where
\begin{displaymath}
A=\frac{3}{4}\frac{1}{L^2-1},\;
B=A+1-\frac{1}{8}\plow,\;
C=\frac{1}{4}(1-\plow)
\end{displaymath}
and 
\begin{displaymath}
D=CN_\rr{K}
\end{displaymath}
The corresponding potassium flux $f^{(1)}_\rr{K}$ given by 
(\ref{flussok}) is depicted in figure~\ref{f:fknew}.
The improvement with respect to $f^{(0)}_\rr{K}$ is striking.

We can further improve the way in which the depletion effect is 
estimated as we did for the sodium particles. 
By neglecting, as we did above, contributions proportional to 
$1/L^2$ and to $1/N_\rr{K}$, 
we get the probability balance equations 
\begin{displaymath}
\left\{
\begin{array}{l}
2q_{\rr{K},0}/4
+
q_{\rr{K}}/4
-
3q_{\rr{K},1}/4
=0
\\
3q_{\rr{K},0}/4
-
q_{\rr{K},2}
+
q_{\rr{K}}/4
=0
\\
q_{\rr{K},2}/4
+
2q_{\rr{K},1}/4
-
3q_{\rr{K}}/4
-
q_{\rr{K}}\plow/8
\\
\;\;\;\;\;\;\;\;\;\;\;\;
\;\;\;\;\;\;\;\;\;
-
(1-\plow)q_\rr{K}(1-r)/4
=0
\\
\end{array}
\right.
\end{displaymath}
and the obvious normalization condition 
\begin{displaymath}
(L^2-4)q_{\rr{K},0}
+
q_{\rr{K},2}
+
2q_{\rr{K},1}
+
q_\rr{K}=1
\end{displaymath} 
The above equations, together with (\ref{erre}),
yields for $r^{(2)}$ and $q^{(2)}_\rr{K}$ 
the same expression as those 
for $r^{(1)}$ and $q^{(1)}_\rr{K}$ 
with 
\begin{displaymath}
A=\frac{25}{48}\frac{1}{L^2-23/12}
\;\textrm{ and }\;
B=\frac{23}{12}A+\frac{37}{48}-\frac{1}{8}\plow
\end{displaymath}
whereas $C$ and $D$ are as before. 
The corresponding potassium flux $f^{(2)}_\rr{K}$, given by 
(\ref{flussok}) and plotted in figure~\ref{f:fknew}, 
improves the estimate $f^{(1)}_\rr{K}$.

We finally comment on the behavior of the occupation 
probability $q_\rr{K}$ as function of the 
low--affinity state probability $\plow$.
In figure~\ref{f:foccu} we have plotted 
$q^{(1)}_\rr{K}$ and $q^{(2)}_\rr{K}$ for $N_\rr{K}=100$ 
and $N_\rr{K}=10000$. We note that 
those curves tend to the same value of the corresponding 
curves for the sodium occupation probability when 
$\plow$ tends to one. This proves that the approximations
are coherent, indeed, in this limit sodium and potassium particles 
behave similarly.
We also note that sodium occupation probabilities are decreasing 
functions of the low--affinity state probability $\plow$.
For potassium ions, the occupation probability decreases at small $\plow$ 
and increases at large $\plow$. This is due to the fact that 
at large $\plow$ the probability $r$ that the pore is occupied 
is small and so is the associated contribution to the potassium flux.

\section{Conclusions}
\label{s:conclusioni}
\par\noindent
Potassium channel are special transmembrane proteins allowing 
selected potassium permeation outside cells. 
Gating and selectivity attitudes are not fully understood. 
For special K$^+$--channel types it has been supposed that 
gating is realized via a selectivity filter. 
In this paper we have proposed and studied a lattice model, in the same 
spirit of~\cite{VanDongen01}, in which a pore selection rule 
ensures both selection and gating. This model has been 
studied both by means of a 
Monte Carlo method and via an analytical approximation.
We have estimated 
both potassium and sodium fluxes through the membrane. 

We have proven the possibility to achieve selection and gating 
and we have discussed the potassium flux reduction due to the 
presence of the selectivity filter. 
In particular we have shown that, allowing more than one potassium 
ion at time to be accommodated inside the pore, the potassium 
flux loss is completely recovered when physiological cytosolic
densities are considered. We then conclude that gating can be 
achieved via a selectivity filter in an efficient way.

We have also studied extensively the model discussing the 
behavior of potassium and sodium fluxes as functions of the 
interesting physiological parameters, i.e., the ionic cytosolic 
density and the time fraction the filter spends in the low--affinity 
state. It is remarkable to notice that the flux of 
the ionic species which is not 
affected by the selectivity filter, sodium in our model, 
is directly proportional to the cytosolic density. 
A sub--linear behavior is found for potassium

In conclusion we think that lattice models provide
essential and tunable tools to investigate efficiently the properties of
ionic channels by analyzing the wealth of experimental data available on
this subject.

\appendix
\section{Detailed definition of the model}
\label{s:bold}
\par\noindent
In this section we describe in detail the Monte Carlo scheme
that we have studied and whose behavior has been 
discussed above.

We consider an integer \textit{time} variable $t$. We set 
$t=0$ and choose at random with uniform probability $1/L^2$ 
the position of the $N_\rr{Na}$ sodium particles and the 
$N_\rr{K}$ potassium  particles.
We then repeat the following \textit{steps} until $t$ equals 
the given integer number $t_\rr{max}$:
\begin{enumerate}
\item
set $t$ equal $t+1$;
\item
select at random the state of the pore: 
choose the low-affinity state with probability 
$\plow$ and the high--affinity one with probability
$1-\plow$;
\begin{enumerate}
\item
if the pore is in the low--affinity state and it is 
occupied by a particle,
the particle is released with the following rule:
it jumps with probability $1/2$ to the site of 
$\Lambda$ neighboring the pore or (with the same probability)
it exits the system;
\item
if a particle exits the system, a particle of the same species is 
put at random with uniform probability $1/L^2$ on one of the $L^2$ sites 
in $\Lambda$;
\end{enumerate}
\item
the position of each particle on the lattice is updated 
following the rules defined in figure~\ref{f:reticolo};
\begin{enumerate}
\item
if a particle enters the pore and the pore is in the 
low--affinity state, the particle is immediately released by the pore 
with the following rule:
it jumps with probability $1/2$ to the site of 
$\Lambda$ neighboring the pore or (with the same probability)
it exits the system;
\item
if a particle exits the system, a particle of the same species is 
put at random with uniform probability $1/L^2$ on one of the $L^2$ sites 
in $\Lambda$.
\end{enumerate}
\end{enumerate}

We considered a second model with the aim of improving the 
potassium flux 
efficiency at low--affinity state probability. 
This model is defined as the basic one with a single 
difference: the rule for a potassium ion in the 
site neighboring the pore when the pore is occupied and in 
the high--affinity 
state (see the right top part of figure~\ref{f:reticolo})
is the same experienced when the pore is free in the high--affinity state.

\section{One dimensional model}
\label{s:uni}
The basic model introduced in section~\ref{s:modello}
can be easily specialized to the one--dimensional case. 
The lattice is $\Lambda=\{1,\dots,L\}$ and the pore 
is the site $L+1$ of $\bb{Z}$.
Equations (\ref{flussona}) and (\ref{flussok}) become
\begin{equation}
\label{flussonauni}
f_\rr{Na}=\frac{1}{4}N_\rr{Na}q_\rr{Na}\plow
\end{equation}
and
\begin{equation}
\label{flussokuni}
f_\rr{K}
=
\frac{1}{4}N_\rr{K}q_\rr{k}\plow
+
\frac{1}{2}\plow r
\end{equation}

In order to compute $q_\rr{Na}$ we consider the random walk of 
a single sodium ion on the lattice $\{1,\dots,L\}$.
The transition matrix elements different from zero are 
\begin{displaymath}
 \pi_{i,i+1}=\frac{1}{2}
\;\textrm{ for } i=1,\dots,L-1
\end{displaymath}
for nearest--neighbor jumps to the right,
\begin{displaymath}
 \pi_{i+1,i}=\frac{1}{2}
\;\textrm{ for } i=1,\dots,L-2
\end{displaymath}
for nearest--neighbor jumps to the left,
\begin{displaymath}
\pi_{L,L-1}=\frac{1}{2}+\frac{1}{4L}\plow
\end{displaymath}
for the nearest--neighbor jump to the left starting from the 
site $L$,
\begin{displaymath}
 \pi_{L,i}=\frac{1}{4L}\plow
\;\textrm{ for } i=1,\dots,L-2
\end{displaymath}
for left jumps starting from the site $L$ (this not zero transition matrix 
elements are due to the fact that particles
exiting the system are put at random with uniform probability 
in one site of the lattice $\Lambda$),
and finally
\begin{displaymath}
\pi_{1,1}
=
\frac{1}{2}
\;\;\textrm{ and }\;\;
\pi_{L,L}
=
\frac{1}{4}\plow
+
\frac{1}{4L}\plow
+
\frac{1}{2}(1-\plow)
\end{displaymath}

We let 
$q_i$, with $i=1,\dots,L$, be the stationary probability 
that the walker occupies the site $i$. Since
the pore is the site $L+1$ of $\bb{Z}$, we have that $q_\rr{Na}$ is nothing 
but $q_L$. 
Recalling the definition of stationary measure
\begin{displaymath}
q_i=\sum_{j=1}^L\pi_{i,j}q_j
\end{displaymath}
for $i=1,\dots,L$, we have 
\begin{equation}
\label{una01}
\begin{array}{rcl}
q_1
&\!\!=&\!\!
{\displaystyle
 \frac{1}{2}q_1
 +
 \frac{1}{2}q_2
 +
 \frac{1}{4L}\plow q_L
 \vphantom{\Bigg[ }
}
\\
q_2
&\!\!=&\!\!
{\displaystyle
 \frac{1}{2}q_1
 +
 \frac{1}{2}q_3
 +
 \frac{1}{4L}\plow q_L
}
\\
&\!\!\vdots&\!\!
\\
q_{L-1}
&\!\!=&\!\!
{\displaystyle
 \frac{1}{2}q_{L-2}
 +
 \frac{1}{2}q_L
 +
 \frac{1}{4L}\plow q_L
 \vphantom{\Bigg[ }
}
\\
q_L
&\!\!=&\!\!
{\displaystyle
 \frac{1}{2}q_{L-1}
 +
 \frac{1}{4}\plow q_L
 +
 \frac{1}{4L}\plow q_L
 +
 \frac{1}{2}(1-\plow) q_L
}
\\
\end{array}
\end{equation}
By using the first $L-1$ equations (\ref{una01}) 
by recursion we get 
\begin{equation}
\label{una02}
q_k
=
q_{k+1}
+
\frac{k}{2L}\plow q_L
\end{equation}
for $k=1,\dots,L-1$.
By recursion, again, it is also easy to prove that 
(\ref{una02}) implies
\begin{equation}
\label{una03}
q_{L-i}
=
q_L
+
\frac{2iL-i(i+1)}{4L}\plow q_L
\end{equation}
for $i=1,\dots,L-1$,
so that the probabilities $q_1,\dots,q_{L-1}$ are all expressed 
in terms of $q_L$. 
Note that $q_{L-i}$ decreases when $i$ runs from $L-1$ to $1$, 
so that the depletion phenomenon discussed in 
section~\ref{s:toy} is found also in the one--dimensional case.

In order to find $q_L$, we finally require that 
the normalization condition
\begin{displaymath}
q_1+\dots+q_L=1
\end{displaymath}
is satisfied.
A straightforward computation yields 
\begin{equation}
\label{una04}
q_L=\frac{1}{L+(2L^2-3L+1)\plow/12}
\end{equation}

\begin{figure}
\begin{picture}(200,200)
\put(-35,170)
{
\includegraphics[height=8.5cm,angle=-90]{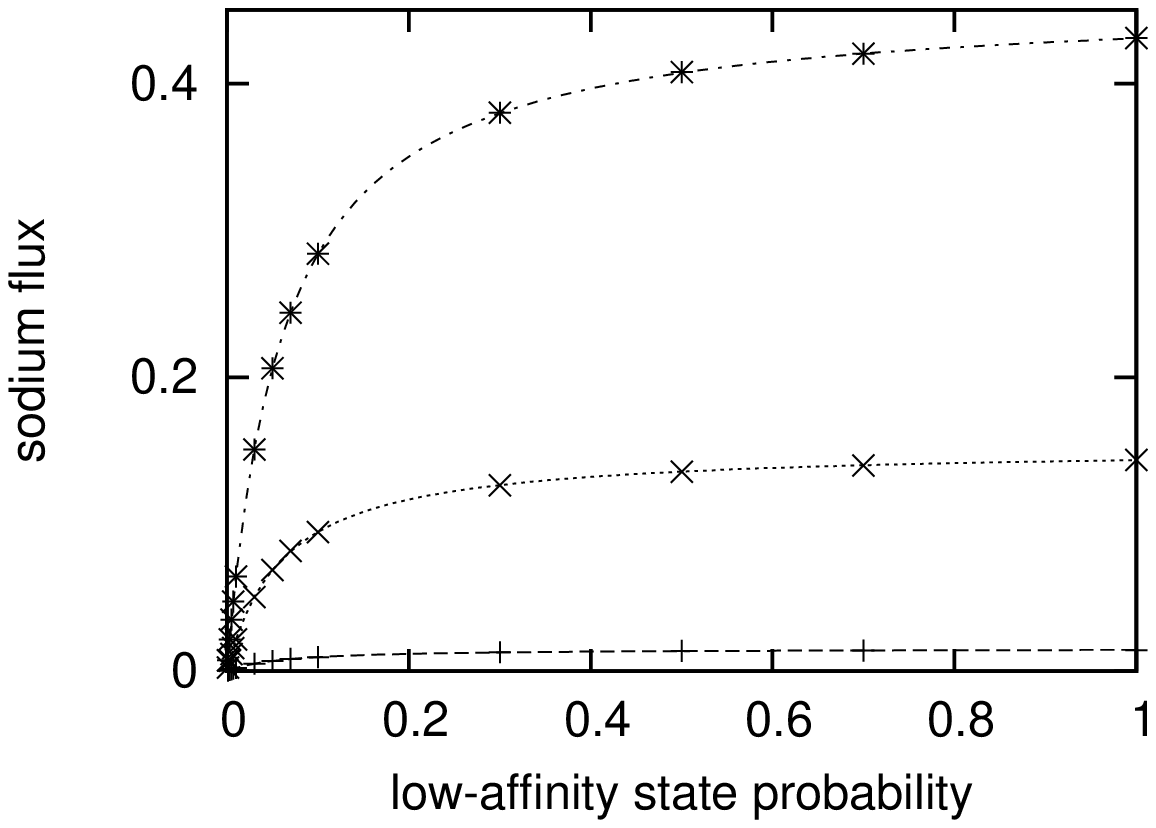}
}
\end{picture}  
\caption{Sodium flux as function of the low--affinity state probability
in the one--dimensional case for $L=100$.
The points $*$, $\times$, and $+$ are the Monte Carlo results 
respectively for $N_\rr{Na}=3000,1000,100$.
The three lines are the graphs of the function (\ref{flussonauni}) 
with $q_\rr{Na}$ given by (\ref{una04}) for the corresponding 
values of $N_\rr{Na}$.
}
\label{f:fnauni} 
\end{figure} 

The one--dimensional sodium flux can be finally computed by 
using (\ref{flussonauni}) with 
$q_\rr{Na}$ given by (\ref{una04}). Monte Carlo results are 
compared with this theoretical prediction in figure~\ref{f:fnauni}
and the matching is perfect.

The computation of $q_\rr{K}$ is more difficult, since the 
interaction between the potassium ions and the pore is not as trivial 
as for sodium. In particular, due to the possibility 
that a potassium ion is trapped in the pore, the potassium walkers 
cannot be considered independent. 
We shall treat this case by assuming that 
in the stationary state the probability for the pore to be 
occupied by one of the $N_\rr{K}$ 
potassium ions is $r$. We then study a potassium 
walker on the lattice $\Lambda=\{1,\dots,L,L+1\}$, where the 
site $L+1$ is the pore, and at the end we shall require that 
$q_{L+1}=r/N_\rr{K}$.
In other words we shall finally assume that, in the stationary state, 
the probability for a single walker to occupy the pore is equal to the 
probability that the pore is occupied by one of the walkers 
divided by the number of potassium ions. 

We do not list all the not zero transition matrix elements, but only those
differing from the sodium case:
\begin{displaymath}
 \pi_{L+1,i}=\frac{1}{2L}\plow
\;\textrm{ for } i=1,\dots,L-1
\end{displaymath}
due to particles released by the pore when it flips to the 
low--affinity state,
and 
\begin{displaymath}
\begin{array}{l}
\pi_{L,L}
=
\plow/4
+
\plow/(4L)
+
(1-\plow)r/2
\\
\pi_{L,L+1}
=
(1-\plow)(1-r)/2
\\
\pi_{L+1,L}
=
\plow/2
+
\plow/(2L)
\\
\pi_{L+1,L+1}
=
(1-\plow)
\end{array}
\end{displaymath}
for the interaction between the pore and its neighboring site.

We repeat, now, the same computation as in the sodium case. We 
first exploit the definition of the stationary measure to write
\begin{equation}
\label{uk01}
\begin{array}{rcl}
q_1
&\!\!=&\!\!
{\displaystyle
 \frac{1}{2}q_1
 +
 \frac{1}{2}q_2
 +
 \frac{1}{4L}\plow q_L
 +
 \frac{1}{2L}\plow q_{L+1}
 \vphantom{\Bigg[ }
}
\\
q_2
&\!\!=&\!\!
{\displaystyle
 \frac{1}{2}q_1
 +
 \frac{1}{2}q_3
 +
 \frac{1}{4L}\plow q_L
 +
 \frac{1}{2L}\plow q_{L+1}
}
\\
&\!\!\vdots&\!\!
\\
q_{L-1}
&\!\!=&\!\!
{\displaystyle
 \frac{1}{2}q_{L-2}
 +
 \frac{1}{2}q_L
 +
 \frac{1}{4L}\plow q_L
 +
 \frac{1}{2L}\plow q_{L+1}
 \vphantom{\Bigg[ }
}
\\
q_L
&\!\!=&\!\!
{\displaystyle
 \frac{1}{2}q_{L-1}
 +
 \frac{1}{4}\plow q_L
 +
 \frac{1}{4L}\plow q_L
 +
 \frac{1}{2}(1-\plow)r q_L
 \vphantom{\Bigg[ }
}
\\
&&\!\!
{\displaystyle
 +
 \frac{1}{2}\plow q_{L+1}
 +
 \frac{1}{2L}\plow q_{L+1}
 \vphantom{\Bigg[ }
}
\\
q_{L+1}
&\!\!=&\!\!
{\displaystyle
 \frac{1}{2}(1-\plow)(1-r) q_L
 +
 (1-\plow) q_{L+1}
}
\\
\end{array}
\end{equation}
As for the sodium we obtain 
\begin{equation}
\label{uk03}
q_{L-i}
=
q_L
+
\frac{2iL-i(i+1)}{4L}\plow q_L
+
\frac{2iL-i(i+1)}{2L}\plow q_{L+1}
\end{equation}
for $i=1,\dots,L-1$,
so that the probabilities $q_1,\dots,q_{L-1}$ are all expressed 
in terms of $q_L$ and $q_{L+1}$. 
By combining (\ref{uk03}) with the expression of $q_L$ 
in (\ref{uk01}) we get 
\begin{displaymath}
q_L=\frac{2\plow}{(1-\plow)(1-r)}q_{L+1}
\end{displaymath}
which is obviously equivalent to the last equation in (\ref{uk01}).

As discussed above, we finally assume $q_{L+1}=r/N_\rr{K}$ and, 
by exploiting the normalization condition 
$q_1+\dots+q_L+q_{L+1}=1$, we get 
\begin{equation}
\label{uk04}
 r=\frac{N_\rr{K}+B+C-\sqrt{(N_\rr{K}+B+C)^2-4N_\rr{K}C}}{2C}
\end{equation}
and
\begin{equation}
\label{uk05}
 q_L=\frac{2\plow}{(1-\plow)(1-r)}\frac{r}{N_\rr{K}}
\end{equation}
where we have set 
\begin{displaymath}
A=\frac{2L^2-3L+1}{3},
B=\frac{2\plow}{1-\plow}\Big(L+\frac{1}{4}A\plow\Big),
C=1+\frac{1}{2}\plow A
\end{displaymath}

\begin{figure}
\begin{picture}(200,200)
\put(-35,170)
{
\includegraphics[height=8.5cm,angle=-90]{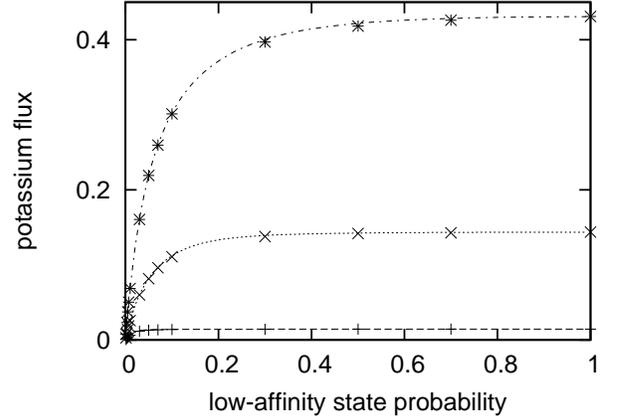}
}
\end{picture}  
\caption{Potassium flux as function of the low--affinity state probability
in the one--dimensional case for $L=100$.
The points $*$, $\times$, and $+$ are the Monte Carlo results 
respectively for $N_\rr{K}=3000,1000,100$.
The three lines are the graphs of the function (\ref{flussokuni}) 
with $q_\rr{K}$ given by (\ref{uk05}) and $r$ given by (\ref{uk04}) 
for the corresponding 
values of $N_\rr{K}$.
}
\label{f:fkuni} 
\end{figure}

The one--dimensional potassium flux can be finally computed by 
using (\ref{flussokuni}) with 
$q_\rr{K}$ given by (\ref{uk05}). Monte Carlo results are 
compared with this theoretical prediction of $r$) in figure~\ref{f:fkuni}
and the matching is perfect.

\begin{figure}
\begin{picture}(200,200)
\put(-35,170)
{
\includegraphics[height=8.5cm,angle=-90]{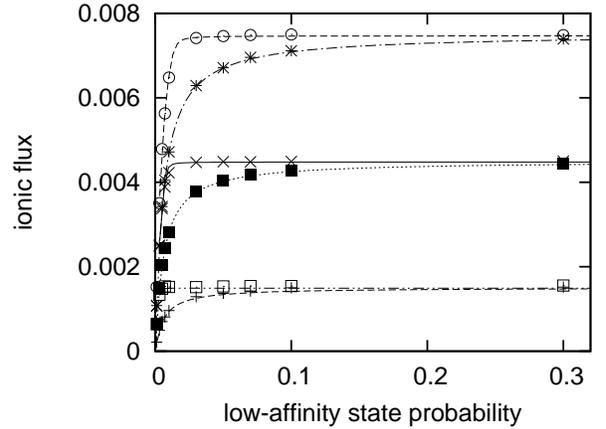}
}
\end{picture}  
\caption{Sodium and potassium flux as function of 
the low--affinity state probability
in the one--dimensional case for $L=1000$.
The points $+$, $\blacksquare$, and $*$ are the Monte Carlo estimates of the 
sodium flux respectively for $N_\rr{Na}=1000,3000,5000$.
The points $\square$, $\times$, and $\circ$ 
are the Monte Carlo estimates of the 
potassium flux respectively for $N_\rr{K}=1000,3000,5000$.
The lines are the graphs of the function 
(\ref{flussonauni}) and (\ref{flussokuni}) 
for the corresponding values of $N_\rr{Na}$ and $N_\rr{K}$.
}
\label{f:fnakuni} 
\end{figure} 

In figure~\ref{f:fnakuni} we have compared Monte Carlo 
and analytical results for $L=1000$. Graphs have been 
plotted on the same figures in order to show that 
even in the one--dimensional case the selection effect can 
be appreciated. 

\begin{acknowledgements}
All numerical simulations have been performed on 
two SUN FIRE 2100 servers of the Dipartimento di Scienze 
di Base e Applicate per l'Ingegneria (SBAI).
We express our thanks to an anonymous refereee
for very useful comments.
\end{acknowledgements}

\end{document}